\documentclass[referee]{aa}
\usepackage{graphicx,natbib}
\usepackage{txfonts}
\usepackage{epic,eepic}

\begin{document}
 
%
%
\def\valid{}    

\font\caps=cmcsc10                  
\font\dunh=cmdunh10  at 12.0 true pt 
\font\dunhs=cmdunh10 
\font\vbold=cmbx10 scaled \magstep1 
\font\sevenbf=cmbx7
\font\sevenit=cmti7
\font\Kapi=cmr17

\def\MEV{DOME}
\def\RTE{equation of radiative transfer}
\def\etal{{et al}}
\def\HW{H\&W}
\def\OK{O\&K}
\def\ok{O\&K}
\def\RH{R\&H}

\def\ibmrs{\hbox{\tt RS/6000}}
\def\hp{\hbox{\tt HP~9000}}
\def\dec{\hbox{\tt DEC~5000}}
\def\axp{\hbox{\tt AXP}}
\def\ibmmf{\hbox{\tt IBM~3090}}
\def\ibmpc{\hbox{\tt 486DX}}
\def\cray{\hbox{\tt Cray 2}}
\def\ymp{\hbox{\tt YMP}}
\def\nec{\hbox{\tt NEC}}

\def\g{\gamma}
\def\b{\beta}
\def\m{\mu}
\def\e{\epsilon}
\def\n{\nu}
\def\l{\lambda}
\def\L{\Lambda}
\def\t{\tau}
\def\pder#1#2{{\partial #1 \over \partial #2}}
\def\div#1#2{{#1\over #2}}
\def\rout{\ifmmode{r_{\rm out}}\else\hbox{$r_{\rm out}$}\fi}
\def\tmax{\ifmmode{\tau_{\rm max}}\else\hbox{$\tau_{\rm max}$}\fi}
\def\tstd{\ifmmode{\tau_{\rm std}}\else\hbox{$\tau_{\rm std}$}\fi}
\def\vmax{\ifmmode{v_{\rm max}}\else\hbox{$v_{\rm max}$}\fi}
\def\muE{\ifmmode{\mu_{\rm E}}\else\hbox{$\mu_{\rm E}$}\fi} 
\def\pE{\ifmmode{p_{\rm E}}\else\hbox{$p_{\rm E}$}\fi} 
\def\bmax{\ifmmode{\b_{\rm max}}\else\hbox{$\b_{\rm max}$}\fi}
\def\kms{\hbox{$\,$km$\,$s$^{-1}$}}
\def\ergs{\hbox{$\,$erg$\,$s$^{-1}$}}
\def\kpc{\hbox{$\,$kpc} }
\def\ang{\hbox{\AA}}
\def\Msun{\hbox{$\,$M$_\odot$} }
\def\Lsun{\hbox{$\,$L$_\odot$} }
\def\Teff{\hbox{$\,T_{\rm eff}$} }
\def\alog#1{\times 10^{#1}}
\def\rin{\hbox{$r_{\rm in}$} }
\def\rout{\hbox{$r_{\rm out}$} }

\def\lstar{\ifmmode{\Lambda^*}\else\hbox{$\Lambda^*$}\fi} 
\def\Rop{\ifmmode{[R_{ij}]}\else\hbox{$[R_{ij}]$}\fi}
\def\Rij{\Rop}
\def\Rji{\ifmmode{[R_{ji}]}\else\hbox{$[R_{ji}]$}\fi}
\def\Rstar{\ifmmode{[R_{ij}^*]}\else\hbox{$[R_{ij}^*]$}\fi}
\def\Rijstar{\Rstar}
\def\Rjistar{\ifmmode{[R_{ji}^*]}\else\hbox{$[R_{ji}^*]$}\fi}
\def\DRji{\ifmmode{[\Delta R_{ji}]}\else\hbox{$[\Delta R_{ji}]$}\fi}
\def\DRij{\ifmmode{[\Delta R_{ij}]}\else\hbox{$[\Delta R_{ij}]$}\fi}

\def\Jb{{\bar J}}
\def\Jnew{{\bar J_{\rm new}}}
\def\Jold{{\bar J_{\rm old}}}
\def\Jfs{{\bar J_{\rm fs}}}
\def\Snew{{S_{\rm new}}}
\def\Sold{{S_{\rm old}}}
\def\Amat{\mat{A}}             

\def\ns{\ifmmode{N_{\rm s}}          
        \else\hbox{$N_{\rm s}$}\fi}
\def\ion#1{\hbox{ #1}}         

\def\peq{\mathbin{\hbox{$+$}\hbox{$=$}}}

\def\mat#1{{\bf #1}}     
\def\vek#1{{#1}}         

\newcount\eqcount
\eqcount=0
\def
  \nummer{
    \global\advance\eqcount by 1
    (\the\eqcount)
  }

\def
  \numadv{
    \global\advance\eqcount by 1
  }

\def
   \numout#1{
     (\the\eqcount #1)
  }

\def\ivek#1#2{\ifmmode{\vek{I}^{#1}_{#2}}
        \else\hbox{$\vek{I}^{#1}_{#2}$}\fi}

\def\ip#1{\ivek{+}{#1}}      
\def\im#1{\ivek{-}{#1}}      

\def\tmat#1#2{\ifmmode{\mat{t}^{#1}_{#2}}
        \else\hbox{$\mat{t}^{#1}_{#2}$}\fi}
\def\rmat#1#2{\ifmmode{\mat{r}^{#1}_{#2}}
        \else\hbox{$\mat{r}^{#1}_{#2}$}\fi}
\def\bvek#1#2{\ifmmode{\beta^{#1}_{#2}}
        \else\hbox{$\beta^{#1}_{#2}$}\fi}

\def\tpi#1{\tmat{+}{#1}}
\def\tmi#1{\tmat{-}{#1}}
\def\rmi#1{\rmat{-}{#1}}
\def\rpi#1{\rmat{+}{#1}}
\def\bpi#1{\bvek{+}{#1}}
\def\bmi#1{\bvek{-}{#1}}

\def\tp{\tmat{+}{}}          
\def\tm{\tmat{-}{}}          
\def\rmm{\rmat{-}{}}         
\def\rp{\rmat{+}{}}          
\def\bp{\bvek{+}{}}          
\def\bm{\bvek{-}{}}          
\def\tpm{\tmat{\pm}{}}       
\def\rpm{\rmat{\pm}{}}       
\def\bpm{\bvek{\pm}{}}       

\def\lp{\ifmmode{\lambda^+_\tau}           
        \else\hbox{$\lambda^+_\tau$}\fi}
\def\lm{\ifmmode\lambda^-_\tau             
        \else\hbox{$\lambda^-_\tau$}\fi}

\baselineskip=12pt

\title{Co-moving frame radiative transfer in spherical media with arbitrary
velocity fields}

\titlerunning{Co-moving frame radiative transfer}
\authorrunning{Baron \& Hauschildt}
\author{E.~Baron\inst{1,2,3} and  Peter H. Hauschildt\inst{4}}

\institute{Dept. of Physics and Astronomy, University of
Oklahoma, 440 W.  Brooks, Rm 131, Norman, OK 73019 USA;
baron@nhn.ou.edu \and
Computational Research Division, Lawrence Berkeley National Laboratory, MS
50F-1650, 1 Cyclotron Rd, Berkeley, CA 94720-8139 USA
\and
Laboratoire de Physique Nucl\'eaire et de Haute Energies, CNRS-IN2P3,
University of Paris VII, Paris, France
\and
Hamburger Sternwarte, Gojenbergsweg 112, 21029 Hamburg, Germany;
yeti@hs.uni-hamburg.de 
}

   \date{Submitted Jan 8, 2004}

\abstract{Recently, with the advances in computational speed and availability
there has been a growth in the number and resolution of fully 3-D
hydrodynamical simulations. However, all of these simulations are
purely hydrodynamical and there has been little attempt to include the
effects of radiative transfer except in a purely phenomenological
manner because the computational cost is too large even for modern
supercomputers. While there has been an effort to develop 3-D Monte
Carlo radiative transfer codes, most of these have been for static
atmospheres or have employed the Sobolev approximation, which limits
their applicability to studying purely geometric effects such as
macroscopic mixing. Also the computational requirements of Monte Carlo methods
are such that it is difficult to couple with 3-D hydrodynamics. Here,
we present an algorithm for calculating 1-D spherical radiative
transfer in the presence of non-monotonic velocity fields in the
co-moving frame. Non-monotonic velocity flows will occur in
convective, and Raleigh--Taylor unstable flows, in flows with multiple
shocks, and in pulsationally unstable stars such as Mira and Cepheids.
This is a first step to developing fully 3-D radiative
transfer than can be coupled with hydrodynamics. We present the
computational method and the results of some test calculations. 
}

\maketitle

\section{Introduction}

The equation of radiative transfer (RTE) in spherical symmetry for
moving media has been solved with a number of different methods, e.g.\
Monte Carlo calculations \citep{magnan70,cns72,avb72}, Sobolev
methods \citep{castor70}, the tangent ray method \citep*{mkh76b}, and
the DOME method \citep{phhrw91}.  Today's state-of-the-art computer codes
use iterative methods for the solution of the RTE, based on the
philosophy of operator splitting or operator perturbation
\citep{cannon73,scharmer84}.  Following these ideas, different approximate
$\Lambda$-operators for this ``accelerated $\Lambda$-iteration'' (ALI)
method have been used successfully \citep*{OAB,hamann87,werner87} and
have been applied to the construction of non-LTE, radiative
equilibrium models of stellar atmospheres
\citep{werner87}. \citet{phhs392} and \citet*{hsb94} have developed an
operator splitting method based on the short-characteristic method
\cite[]{OAB,ok87} to obtain the formal solution of the special
relativistic, spherically symmetric RTE along its characteristics and
a band-diagonal approximation to the discretized
$\Lambda$-operator. This method can be implemented very efficiently to
obtain an accurate solution of the spherically symmetric RTE for
continuum and line 
transfer problems using only modest amounts of computer resources.

The main restriction of the co-moving frame (CMF) method discussed in
\citet{phhs392} is a restriction to monotonic velocity fields. For
monotonic velocity fields, the wavelength derivative part of the CMF
RTE can be posed as an initial value problem with the initial
conditions set at small wavelengths for expanding media and at long
wavelengths for contracting media. The initial value problem has to be
solved by a fully implicit wavelength discretization (e.g., upwind
schemes) in order to guarantee stability. In media with non-monotonic
velocity fields, the wavelength derivative changes the structure of
the equation so that it becomes a boundary value
problem with boundary conditions at both short and long wavelengths.
Since the equation is first order in wavelength at each
spatial point there is only one boundary condition, whose wavelength
sense depends on the local sign of the coefficient of the
derivative. Therefore, if the wavelength-space is discretized to solve
the CMF RTE, we need {\em local} upwind schemes in order to guarantee
stability and to properly account for the presence of mixed boundary
conditions in wavelength-space.

In principle, it is possible to solve the RTE in the observer's frame,
however, in that frame the emission and absorption processes are
anisotropic and a detailed calculation is extremely complex.  The main
obstacle at high velocity, is that a prohibitively large number of
wavelength and angle points are required to solve the observer's
frame RTE.

In this paper, we describe an operator splitting method to solve the
spherical relativistic CMF RTE for arbitrary velocity fields. The
method can be applied to a wide variety of astrophysical problems;
e.g., atmospheres of pulsating stars, stellar atmospheres with shocks,
multi-component novae, and supernova atmospheres. Although we present
the method for the spherically symmetric 1-dimensional case, it can be
extended to 3D geometry. Our approach calculates the formal solution
along each characteristic independently (for known source function) in
the combined radius-wavelength space.  This has to be done in order to
account for the wavelength and radial boundary conditions
simultaneously  during the formal solution stage of the
operator splitting method. As the ``approximate $\L$ operator'' we
use a block matrix that is tri-diagonal in wavelength space where each
spatial (i.e., radial) block is itself a band matrix. This
``wavelength tri-diagonal'' approach yields superior convergence
compared to a simpler ``wavelength diagonal'' approach (which would
correspond to a diagonal approximate $\L$ operator in static radiative
transfer problems). The work presented here is a first step in an
effort to develop methods for solving the CMF RTE in fully 3-D
geometry, but even in its spherical version presented here will be
useful for studying varying stars, such as Miras and Cepheids.

\section{Method}

In the following discussion we use  notation of \citet{phhs392}.
Our starting point is  the spherically symmetric form of the special
relativistic,  
time independent ($\partial/\partial t \equiv 0$) RTE, the restriction 
to plane parallel geometry is straightforward. The calculation of
the characteristics is identical to \citet{phhs392} and we thus assume that
the characteristics are known.
First, we will describe the process for the formal solution, then we will
describe how we construct the approximate $\L$ operator, $\lstar$.

\subsection{Formal solution}

We begin with Eq.~16 of \citet{phhs392} \citep[see also Eq.~1 of][]{hb04}:
\begin{equation}
\frac{dI_l}{ds} + a_l\frac{\partial \l I}{\partial \l} = \eta_l - 
(\chi_l+4a_l)I_l \label{rte-char}
\end{equation}
where $ds$ is a line element along a (curved) characteristic, $I_l(s)$ 
is the specific intensity along the characteristic at point $s\ge 0$ ($s=0$
denotes the beginning of the characteristic) and wavelength point $\l_l$.
The coefficient $a_l$ is defined by
\[
     a_l = \gamma  
                   \left[ \frac{\beta(1-\mu^2)}{r}
                         +\gamma^2\mu\left(\mu+\beta\right)\pder{\beta}{r}
                   \right]                        
\]
where $\beta=v/c$, $\gamma=\sqrt{1-\beta^2}$ and $r$ is the radius.
$\eta_l$ and $\chi_l$ are the emission and extinction coefficients 
at wavelength $\l_l$, respectively.


\def\plm{p_{l,l-1}}
\def\plp{p_{l,l+1}}
\def\pll{p_{l,l}}
\def\Ilm{I_{l-1}}
\def\Ilp{I_{l+1}}
\def\Ill{I_{l}}

Equation \ref{rte-char} describes the change of the intensity along a arbitrary
characteristic though the medium. 
We define
\[
\hat\chi_l \equiv \chi_l+4a_l
\]
and discretize the wavelength derivative
using a 3 point differencing formula (this can be made more general) to obtain:
\begin{equation}
\label{rte-discr}
\frac{dI_l}{ds} + a_l \left[ \plm\Ilm + \pll\Ill + \plp\Ilp \right]
= \eta_l -(\chi_l+4a_l)\Ill
\end{equation}
where $S_l=\eta_l/\chi_l$ is the source function at $\l=\l_l$, 
$d\tau  = \hat\chi\,ds$, and  the $p_{ij}$ are the discretization
coefficients for the  
$\partial \l I_l/\partial \l$ derivatives (see
Eqns~\ref{didlam_expdng}--\ref{didlam_cntrctng} for details). 

To obtain an expression for the formal solution, we rewrite Eq.~\ref{rte-discr}
as
\[
\frac{dI_l}{d\tau} = \Ill - \hat S_l - \tilde S_l 
\]
with
\begin{eqnarray*}
\hat S_l &= &\frac{\chi_l}{\hat\chi_l}S_l = \frac{\eta_l}{\hat\chi_l} \\
\tilde S_l &=
&-\frac{a_l}{\hat\chi_l}\left[\plm\Ilm+\pll\Ill+\plp\Ilp\right] \\ 
\end{eqnarray*}
(note that the index $l$ denotes the wavelength point $\l_l$)


\def\Imm{I_{i-1,l-1}}
\def\Imp{I_{i-1,l+1}}
\def\Ipm{I_{i+1,l-1}}
\def\Ipp{I_{i+1,l+1}}
\def\Iil{I_{i,l}}
\def\Iml{I_{i-1,l}}

\def\mm{_{i-1,l-1}}
\def\mp{_{i-1,l+1}}
\def\ml{_{i-1,l}}
\def\pm{_{i+1,l-1}}
\def\pp{_{i+1,l+1}}
\def\il{_{i,l}}
\def\im{_{i,l-1}}
\def\ip{_{i,l+1}}
\def\ml{_{i-1,l}}
\def\pl{_{i+1,l}}

With this we obtain the following expression for
the formal solution \citep[see also Eq.~14 in][]{hb04}
\begin{equation}
\Iil = \Iml\exp(-\Delta\tau_{i-1}) + \delta\hat\Iil + \delta\tilde\Iil
\label{fs}
\end{equation}
with the definitions
\[
\delta\hat\Iil = \alpha\il \hat S\ml + \beta\il \hat S\il + \gamma\il \hat S\pl
\]
and
\[
\delta\tilde\Iil = \alpha\il \tilde S\ml + \beta\il \tilde S\il 
\]
The index $i$ labels the (spatial) points along a characteristic, the index
$l$ denotes the wavelength point. The coefficients $\alpha\il$, $\beta\il$,
and $\gamma\il$ are given in \citet{phhs392} and \citet{ok87}, here they are 
calculated for a fixed wavelength for all points along a characteristic.
$\hat S$ is a vector of known quantities (the old mean intensities and 
thermal sources). The $\tilde S$ contain the effects of the velocity field
on the formal solution and are given by
\begin{eqnarray*}
\tilde S\ml & = & -\frac{a\ml}{\hat\chi\ml} 
\left[ \plm I\mm +\pll I\ml + \plp I\mp \right]\\
\tilde S\il & = & -\frac{a\il}{\hat\chi\il} 
\left[ \plm I\im + \pll \Iil+ \plp I\ip \right]\\
\end{eqnarray*}
This term provides the coupling of different wavelengths in the formal
solution. Note that we have only used linear interpolation of the
$\tilde S$ terms and that this differencing scheme is different than
that of \citet{phhs392}. We have described this differencing scheme in
detail in \citet{hb04}. We have found that this differencing scheme is
less diffusive in the co-moving frame and thus produces better line
profiles for the case of small differential expansion. It gives
identical results in the case of large global differential expansion
(novae and supernovae) where the numerical diffusion in the co-moving
frame is overwhelmed by the global line width. We have shown that this
differencing scheme can become unstable under certain conditions, but
can be stabilized in a straightforward manner \citep{hb04}. We present
here the more complex $\zeta=1.0$, the generalization of the
differencing scheme to the fully stable differencing scheme is obvious
and can be found in \citet{hb04}.

If the velocity field is
monotonically increasing or decreasing, then $\plp\equiv 0$ or
$\plm\equiv 0$ for a stable upwind discretization of the wavelength
derivative. In these cases, the problem becomes an initial value
problem and can be solved for each wavelength once the results of the
previous (smaller or longer) wavelength points are known.  For
non-monotonic velocity fields this is no longer the case and the
formal solution needs to explicitly account for the wavelength
couplings in both the blue and red directions.

The formal solution is equivalent to the solution of one linear system
for each characteristic. The rank of the system is $n_l\times n_i$
where $n_l$ is the number of wavelength points and $n_i$ is the number
of points along the characteristic. For $n_r$ radial points, we have
$3 \le n_i \le 2n_r-1$ points along each characteristic. The number of
wavelength points $n_l$ can be much larger, $n_l\approx 1000$ for the
test cases presented later in this paper but $n_l \approx 300000$ in
full scale applications, so the rank of the systems can become
large. Fortunately, the linear systems have a simple structure that
allows us to use efficient methods for their numerical solution.

For the construction of the system matrix for the formal solution along
each characteristic it is useful to write the coefficients
$k_{i,l}$ of the intensities as follows:
\[ 
\begin{array}{llcl}
I\mm: &  k\mm & = & \displaystyle -\frac{\alpha\il a\ml \plm}{\hat \chi\ml}\\[15pt]
I\ml: &  k\ml & = & \displaystyle \exp(-\Delta\tau_{i-1}) \\[15pt]
I\mp: &  k\mp & = & \displaystyle -\frac{\alpha\il a\ml \plp}{\hat \chi\ml}\\[15pt]
I\im: &  k\im & = & \displaystyle -\frac{\beta\il a\il \plm}{\hat \chi\il}\\[15pt]
I\il: &  k\il & = & \displaystyle -\frac{\beta\il a\il \pll}{\hat \chi\il}\\[15pt]
I\ip: &  k\ip & = & \displaystyle -\frac{\beta\il a\il \plp}{\hat \chi\il}\\[15pt]
I\pm: &  k\pm & = & \displaystyle -\frac{\gamma\il a\pl \plm}{\hat \chi\pl}\\[15pt]
I\pp: &  k\pp & = & \displaystyle -\frac{\gamma\il a\pl \plp}{\hat \chi\pl}\\[15pt]
\end{array}
\]
These expressions show the relatively simple matrix structure that can
be exploited to solve 
for the mean intensities.

\subsubsection{Discretization of $\partial \l I / \partial \l$}

In order to ensure numerical stability, we use a local upwind scheme to
discretize the wavelength derivative in the RTE:  
\begin{itemize}
\item For $a_l \ge 0$:
\begin{eqnarray}
\left.\frac{\partial \l I}{\partial \l}\right|_l =
\frac{\l_l I_l - \l_{l-1} I_{l-1}}{\l_l-\l_{l-1}}
\label{didlam_expdng}
\end{eqnarray}
\item For $a_l < 0$:
\begin{eqnarray}
\left.\frac{\partial \l I}{\partial \l}\right|_l =
\frac{\l_l I_l - \l_{l+1} I_{l+1}}{\l_l-\l_{l+1}}
\label{didlam_cntrctng}
\end{eqnarray}
\end{itemize}
Here, and in the following, we use a sorted wavelength grid with
$\l_{l-1}<\l_l<\l_{l+1}$. The wavelength derivative is evaluated at a
fixed spatial point along the characteristic. The coefficients $p$ are
then given by 
\begin{itemize}
\item For $a_l \ge 0$:
\begin{eqnarray*}
\plm & = & -\frac{\l_{l-1}}{\l_l-\l_{l-1}} \\
\pll & = & \frac{\l_l}{\l_l-\l_{l-1}} \\
\plp & = & 0 \\
\end{eqnarray*}
\item For $a_l < 0$:
\begin{eqnarray*}
\plm & = & 0 \\
\pll & = & \frac{\l_l}{\l_l-\l_{l+1}} \\
\plp & = & -\frac{\l_{l+1}}{\l_l-\l_{l+1}} \\
\end{eqnarray*}
\end{itemize}
These coefficients depend on both the radial coordinate (because $a_l$ is
a function of $r$) and on the wavelengths (in the general case of a
wavelength grid with variable resolution). Note that the
\emph{direction} of flow of information is determined by the sign of
the coefficient $a_l$ and not just on the velocity gradient. 

\subsubsection{Boundary Conditions}

The spatial boundary conditions in the non-monotonic case remain the
same as in the monotonic case, the incoming intensities at the spatial
boundaries of each characteristic must be prescribed at every
\emph{wavelength} point.

The wavelength boundary conditions are a bit more complicated, since
now at every spatial point there is a wavelength boundary condition
which must be determined by the local flow of information which is
determined by the sign of each $a_l$ along the characteristic. I.e.,
at each spatial point, one must determine whether information is
flowing from blue-to-red or red-to-blue (in wavelength) and implement
the proper 
boundary condition.

\subsubsection{Structure of the system matrix}

We label the total number of wavelength points $n_l$, and the 
number of intersection points along a characteristic \citep[see][]{hsb94}
with $n_i$. The total number of intensities that need to be determined
is thus $n_i \times n_l$ per characteristic. Note that $n_i$ depends on the 
characteristic that is used in spherical symmetry and in the general
3-D case. There are two different ways to write the 
vector $\vec I$ of the specific intensities:
\begin{enumerate}
\item ``i-ordering'': $\vec I = \vec {\left(\vec I_i\right)_l}$, so that
$\vec I$ is a vector of $n_l$ vectors each of which has $n_i$ components.
\item ``l-ordering'': $\vec I = \vec {\left(\vec I_l\right)_i}$, so that
$\vec I$ is a vector of $n_i$ vectors each of which has $n_l$ components.
\end{enumerate}
The position of the element $(i,l)$ is therefore:
\begin{enumerate}
\item ``i-ordering'': $I_{il} = (l-1)n_i+i$
and
\item ``l-ordering'': $I_{li} = (i-1)n_l+l$.
\end{enumerate}
In the following, we will denote these ``block-vectors'' with $(n_i,n_l)$ 
for both ordering schemes. 

If we write Eq.~\ref{fs} in matrix form, we obtain
\[
\vec I = A \vec I + \hat S + \tilde S
\]
where $\hat S$ is a $(n_i,n_l)$ vector with the thermal and scattering
source functions for each wavelength point $n_l$ and radial point
$n_i$ functions for each wavelength point, $\tilde S$ is a $(n_i,n_l)$
vector with the wavelength derivative information, and $A$ is a
$(n_i,n_l)\times (n_i,n_l)$ matrix. The row $(i,l)$ of $A$ has the
following entries (the location of the element for i-ordering is also
given):
\[
\begin{array}{lll}
\rm location & \rm index  & \rm matrix\ element \\
 (i-1,l-1)  &  (l-2)n_i+i-1  &  k\mm \\
 (i-1,l+1)  &  l n_i+i-1     &  k\mp \\
 (i,l-1)    &  (l-2)n_i+i    &  k\im \\
 (i-1,l)      &  (l-1)n_i+i-1    &  k\ml \\
 (i,l)      &  (l-1)n_i+i    &  k\il \\
 (i,l+1)    &  l n_i+i       &  k\ip \\
 (i+1,l-1)  &  (l-2)n_i+i+1  &  k\pm \\
 (i+1,l+1)  &  l n_i+i+1     &  k\pp \\
\end{array}
\]
The total bandwidth of $A$ in the i-ordering scheme is $l
n_i + i+1 -((l-2)n_i+i-1) = 2(n_i+1)$. The l-ordering scheme is
symmetric to the i-ordering, thus the bandwidth in l-ordering is
simply $2(n_l+1)$. For large $n_l$, i-ordering will require much
smaller bandwidth, therefore, we consider only the i-ordering scheme
here.

\newcommand{\DD}{\ensuremath{B^{diag}}}
\newcommand{\DSB}{\ensuremath{A^{diag}}}
\newcommand{\SPD}{\ensuremath{B^{super}}}
\newcommand{\SPSB}{\ensuremath{A^{super}}}
\newcommand{\SPSP}{\ensuremath{C^{super}}}
\newcommand{\SBD}{\ensuremath{B^{sub}}}
\newcommand{\SBSB}{\ensuremath{A^{sub}}}
\newcommand{\SBSP}{\ensuremath{C^{sub}}}

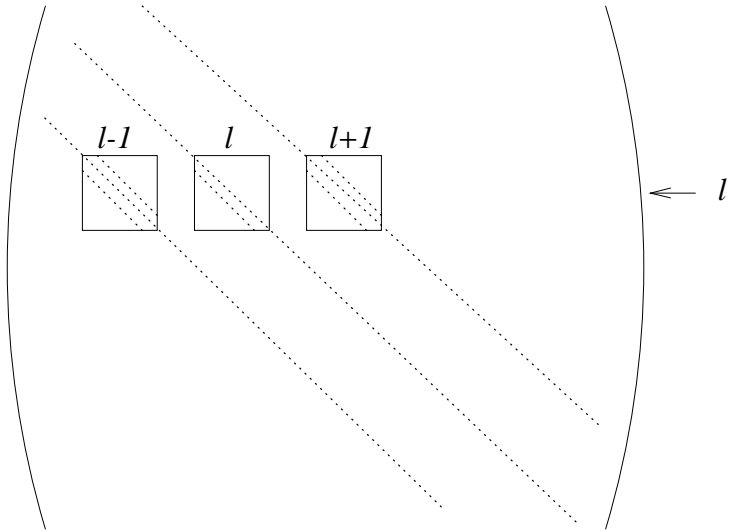
\begin{figure}[hbt]
\centering
\setlength{\unitlength}{0.00087489in}
\begingroup\makeatletter\ifx\SetFigFont\undefined%
\gdef\SetFigFont#1#2#3#4#5{%
  \reset@font\fontsize{#1}{#2pt}%
  \fontfamily{#3}\fontseries{#4}\fontshape{#5}%
  \selectfont}%
\fi\endgroup%
{\renewcommand{\dashlinestretch}{30}
\begin{picture}(4335,3185)(0,-10)
\path(464,2258)(914,2258)(914,1808)
	(464,1808)(464,2258)
\path(1139,2258)(1589,2258)(1589,1808)
	(1139,1808)(1139,2258)
\path(1814,2258)(2264,2258)(2264,1808)
	(1814,1808)(1814,2258)
\put(5526.500,1583.000){\arc{11034.180}{2.8521}{3.4311}}
\put(-1673.500,1583.000){\arc{11034.180}{5.9937}{6.5727}}
\dottedline{45}(239,2483)(2624,143)
\dottedline{45}(824,3158)(3569,638)
\path(3884,2033)(4154,2033)
\path(4004.000,2063.000)(3884.000,2033.000)(4004.000,2003.000)
\dottedline{45}(419,2933)(3434,53)
\dottedline{45}(464,2168)(824,1808)
\dottedline{45}(554,2258)(914,1898)
\dottedline{45}(1814,2168)(2174,1808)
\dottedline{45}(1904,2258)(2264,1898)
\dottedline{45}(1139,2168)(1499,1808)
\put(1319,2303){\makebox(0,0)[lb]{\smash{{{\SetFigFont{12}{14.4}{\rmdefault}{\mddefault}{\itdefault}l}}}}}
\put(554,2303){\makebox(0,0)[lb]{\smash{{{\SetFigFont{12}{14.4}{\rmdefault}{\mddefault}{\itdefault}l-1}}}}}
\put(1949,2303){\makebox(0,0)[lb]{\smash{{{\SetFigFont{12}{14.4}{\rmdefault}{\mddefault}{\itdefault}l+1}}}}}
\put(4289,1988){\makebox(0,0)[lb]{\smash{{{\SetFigFont{12}{14.4}{\rmdefault}{\mddefault}{\itdefault}l}}}}}
\end{picture}
}

\caption{\label{blockmatrix} Structure of the system matrix.
}
\end{figure}

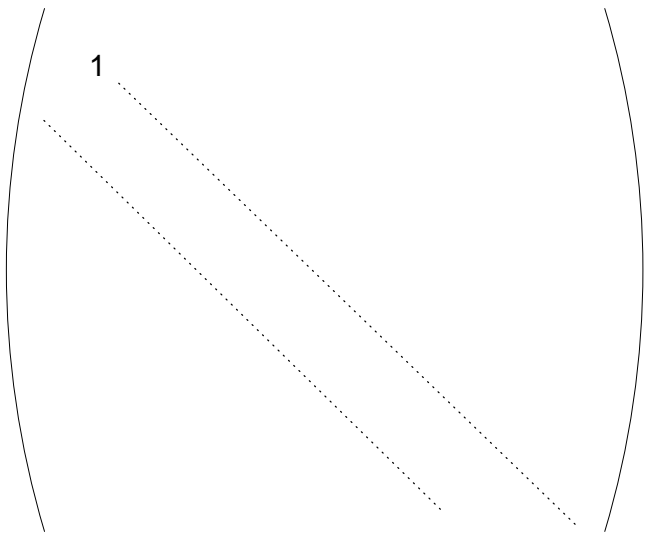
\begin{figure}[hbt]
\centering
\setlength{\unitlength}{0.00087489in}
\begingroup\makeatletter\ifx\SetFigFont\undefined%
\gdef\SetFigFont#1#2#3#4#5{%
  \reset@font\fontsize{#1}{#2pt}%
  \fontfamily{#3}\fontseries{#4}\fontshape{#5}%
  \selectfont}%
\fi\endgroup%
{\renewcommand{\dashlinestretch}{30}
\begin{picture}(3852,3181)(0,-10)
\put(5526.500,1583.000){\arc{11034.180}{2.8521}{3.4311}}
\put(-1673.500,1583.000){\arc{11034.180}{5.9937}{6.5727}}
\dottedline{45}(239,2483)(2624,143)
\dottedline{45}(689,2708)(3434,53)
\put(509,2753){\makebox(0,0)[lb]{\smash{{{\SetFigFont{12}{14.4}{\sfdefault}{\mddefault}{\updefault}1}}}}}
\end{picture}
}

\caption{\label{l-block} Structure of the $l$ block.
}
\end{figure}

To calculate the intensities along each characteristic we have to solve 
the linear system
\begin{equation}
(1-A) \vec I = \hat S + \tilde S \label{system}
\end{equation}
for each characteristic where $I$ is the identity matrix.  In the
i-ordering scheme, the system matrix ${\cal A} = I - A$ is thus a
block-tridiagonal matrix with $n_l$ blocks of $n_i \times n_i$
matrices. Fig.~\ref{blockmatrix} shows the general structure of the
matrix ${\cal A}$. The
$l-1$ and $l+1$ blocks are tridiagonal matrices, the $l$-block has the
layout shown in Fig.~\ref{l-block}. Note the sub-diagonal in the
$l$-block is just $-\exp(-\Delta\tau_{i-1})$. For the static case, the
system degenerates to one with just the $l$-block non-zero 
allowing for direct recursive solution of the problem. For monotonic
velocity fields either the $l-1$ or the $l+1$ blocks are zero, 
again admitting recursive solution. In the case of non-monotonic
velocity fields, all blocks may be non-zero and the system must be
solved explicitly. It is convenient to call the tridiagonal matrix
labeled $l+1$ in
Figure~\ref{blockmatrix} $super$, the lower diagonal matrix labeled
$l$, $diag$, and the tridiagonal matrix labeled $l-1$, $sub$. Then we
can refer to e.g., the lower diagonal, diagonal, and upper diagonal of
$super$ as \SPSB, \SPD, and \SPSP, respectively.

\subsubsection{Solution of the linear systems}

The linear system of Eq.~\ref{system} has a relatively simple
structure and can be solved directly, e.g., by block tri-diagonal
system solvers as described in \citet{golub89:_matrix}. The problem with
this approach is that the inverse of a tri-diagonal matrix is, in
general, a {\em full} matrix. Therefore, the CPU time and memory
requirements for direct solvers increase dramatically with increasing
$n_l$.  The special form of Eq.~\ref{system} and the sparseness of the
blocks within the matrix $A$ led us to investigate iterative methods for the
solution of Eq.~\ref{system}. We examined the use of the ``Rapido'' algorithm
\citep[][page 194ff]{matrizen}, however, as with all iterative linear
system solvers, the eigenvalues of the matrix
can (and are in many cases) be such that this method fails. For this
work we also used 
standard band matrix solvers available in LAPACK and ESSL, which have
the advantage that the pivots only have to be calculated once per ALO
iteration. Finally we used the general sparse solver package
SuperLU \citep{super_lu03} which leads to large speedups over the
other linear systems solvers we have tried (see below for a discussion
of timing).

\subsection{Construction of $\lstar$}

The major difference between the monotonic velocity field problem and
the non-monotonic velocity field is that
we now have to construct a combined spatial-wavelength approximate
$\L$ operator for use in the operator splitting scheme. The basic
equations of the operator splitting method remain unchanged,
however. In the discussion of the construction of $\lstar$ it is
useful to consider the ``spatial'' part of the $\L$ matrix and the
``wavelength'' part of the $\L$ matrix.  The spatial part of the $\L$
matrix describes the transfer of photons in the radial coordinate and
is, essentially, identical to the monotonic velocity field problem. The
wavelength part of the $\L$ matrix describes the transport of photons
in wavelength space due to the velocity field (or other non-coherent
scattering processes).

The $\Lambda$ operator at wavelength point $l$, $\Lambda_l$,
has contributions from {\em all} wavelength points.  This would lead
to a matrix of order $n_l\times n_r$, where $n_r$ denotes the number
of radial points. With a band-matrix form for the spatial $\lstar$'s,
this would lead to a global band-matrix with significant storage
requirements.  Therefore, we derive a tri-diagonal approximation to
the wavelength contributions to the global $\lstar$ matrix.

The construction of the {\em spatial} part of $\lstar$ proceeds in
exactly the same way as described in \citet{hsb94}, that is we assume
a pulse of intensity of value unity is inserted into the
characteristic and by acting the $\L$ matrix on the pulse we are able
to construct and approximate lambda operator
\citep[ALO][]{OAB,ok87}. Even though we have 
made use of the fact that the formal solution can be written as a
tridiagonal operator we shall show that each component of the ALO
contains effects of both spatial and wavelength propagation of the
global $\L$ matrix.

We describe the  construction of $\lstar$ for arbitrary (spatial)
bandwidth using the example of a tangential ray (core 
intersecting rays are a simple specialization of this case): 
The 
intersection points (including the point of tangency) are labeled from left 
to right, the direction in which the formal solution proceeds. 
For convenience, we label the ray tangent to shell $i+1$ as $i$. Therefore, the
ray $i$ has $2i+1$ points of intersection with discrete shells
$1\ldots i+1$. For each point $k$ along the ray there is a ``mirror
point'' $k_m = 2i + 1 - k$.
To compute 
row $j$ of the discrete 
$\Lambda$-operator (or $\Lambda$-matrix), $\Lambda_{ij}$, we sequentially 
label the 
intersection points of the ray $i$ with the shell $j$ (``running 
index''), 
and define auxiliary quantities $\xi_{k,l}^i$. The pulse is
inserted at point $k_s$, which is either $k-1$ or point $k=0$ in the
case $j=1$. 
It is convenient to define an $X$-factor as follows:
\[
X = \left\{ \begin{array}{ll}
\frac{\chi_{k_s,l}}{\hat\chi_{k_s,l}}\beta_{k_s,l}&\mbox{if $k=0$}\\
\frac{\chi_{k_s,l}}{\hat\chi_{k_s,l}}\gamma_{k_s,l}&\mbox{if $k>0$}\\
	    \end{array}
\right.
\]
Then 
\[ \xi^i_{k_s,l} = \frac{X}{1-\DD_{k_s,l}} \]
\[ \xi^i_{k_s,l-1} = \frac{\SPD_{k_s,l-1}\xi^i_{k_s,l}}{1-\DD_{k_s,l-1}} \]
\[ \xi^i_{k_s,l+1} =
\frac{\SBD_{k_s,l+1}\xi^i_{k_s,l}}{1-\DD_{k_s,l+1}} \]
Then, propagating the pulse through the grid we obtain
\[
X_j = \left\{ \begin{array}{ll}
\frac{\chi_{j,l}}{\hat\chi_{j,l}}\beta_{j,l}&\mbox{if $j=k$}\\
\frac{\chi_{j,l}}{\hat\chi_{j,l}}\alpha_{j,l}&\mbox{if $j=k+1$}\\
X_j + \frac{\chi_{j,l}}{\hat\chi_{j,l}}\gamma_{j,l}&\mbox{if $j=k_m-1$}\\
\frac{\chi_{j,l}}{\hat\chi_{j,l}}\beta_{j,l}&\mbox{if $j=k_m$}\\
\frac{\chi_{j,l}}{\hat\chi_{j,l}}\alpha_{j,l}&\mbox{if $j=k_m+1$}\\
	    \end{array}
\right.
\]

Then for $a_{j,l} < 0$ we have
\begin{eqnarray}
\xi^i_{j,l+1} &=&
({1-\DD_{j,l+1}})^{-1}[\SBSB_{j,l+1}\xi^i_{j-1,l} +
\DSB_{j-1,l+1}\xi^i_{j-1,l+1} ] \nonumber\\
\xi^i_{j,l} &=& ({1-\DD_{j,l}})^{-1}[X +
\SBSB_{j,l}\xi^i_{j-1,l-1} + \DSB_{j,l}\xi^i_{j-1,l} +
\SPSB_{j,l}\xi^i_{j-1,l+1} + \SPD_{j,l}\xi^i_{j,l+1} ] \nonumber\\
\xi^i_{j,l-1} &=&
({1-\DD_{j,l-1}})^{-1}[\DSB_{j,l-1}\xi^i_{j-1,l-1} + 
\SPSB_{j,l-1}\xi^i_{j-1,l} + \SPD_{j,l-1}\xi^i_{j,l} ] \nonumber
\end{eqnarray}
and for $a_{j,l} \ge 0$ we have
\begin{eqnarray}
\xi^i_{j,l-1} &=&
({1-\DD_{j,l-1}})^{-1}[\DSB_{j,l-1}\xi^i_{j-1,l-1} + 
\SPSB_{j,l-1}\xi^i_{j-1,l}  ] \nonumber\\ 
\xi^i_{j,l} &=& ({1-\DD_{j,l}})^{-1}[X +
\SBSB_{j,l}\xi^i_{j-1,l-1} + \DSB_{j,l}\xi^i_{j-1,l} +
\SPSB_{j,l}\xi^i_{j-1,l+1} + \SPD_{j,l}\xi^i_{j,l+1} ] \nonumber\\ 
\xi^i_{j,l+1} &=&
({1-\DD_{j,l+1}})^{-1}[\SBSB_{j,l+1}\xi^i_{j-1,l} +
\DSB_{j-1,l+1}\xi^i_{j-1,l+1}  + \SBD_{j,l+1}\xi^i_{j,l}] \nonumber
\end{eqnarray}

With this we can
construct  $\lstar$'s with the full spatial bandwidth and
tridiagonal in wavelength.  However, in
order to obtain good convergence, we also have to include wavelength
dependent information in the global (spatial plus wavelength)
$\lstar$.  

We can write the $l$ component of the formal solution $\Jfs$ in the
form
\begin{equation}
(\Jfs)_l = (\Lambda S)_l \approx \Lambda_{l,l-1} S_{l-1} 
                   + \Lambda_{l,l} S_{l}
                   + \Lambda_{l,l+1} S_{l+1}\label{alo}
\end{equation}
Here, $\Lambda_{l,l-1}$, $\Lambda_{l,l}$, and $\Lambda_{l,l+1}$ are
the contributions to the $\Lambda$ operator at wavelength
point $l$ originating from the wavelength points $l-1$, $l$, and
$l+1$.  These matrices can be computed directly from the
$\lambda^i_{j,l}$ calculated above, simply by integrating over angle.
Equation \ref{alo} shows
explicitly the dependence of 
the mean intensities on the source functions of the neighboring
wavelength points. Therefore, we can construct a global $\lstar$
operator with the definition
\[
(\lstar S)_l \equiv \lstar_{l,l-1} S_{l-1} 
                   + \lstar_{l,l} S_{l}
                   + \lstar_{l,l+1} S_{l+1}
\]
This leads to a block tri-diagonal global $\lstar$, where each spatial
block is again a band matrix with the band-width given by the full
spatial $\lstar$.  This system can be solved efficiently either by
direct solvers, using the same methods as discussed
above for the formal solution. The convergence of the operator
splitting method using the above $\lstar$ is similar to that of static
operator splitting methods with a (spatial) tri-diagonal
ALO. Typically, 15-20 operator splitting iterations are required to
reach an accuracy of $10^{-8}$ for test cases with about 1000 wavelength
points (see below). A wavelength-diagonal operator converges much more
slowly, and an operator that ignores the wavelength dependence would
require a maximum of $n_l\times n_{\rm iter}$ iterations where $n_{\rm
iter}$ is the number of iterations required to solve a monotonic
velocity field problem
for a single wavelength.

\begin{figure}[hbt]
\centering
\setlength{\unitlength}{0.00087489in}
\begingroup\makeatletter\ifx\SetFigFont\undefined%
\gdef\SetFigFont#1#2#3#4#5{%
  \reset@font\fontsize{#1}{#2pt}%
  \fontfamily{#3}\fontseries{#4}\fontshape{#5}%
  \selectfont}%
\fi\endgroup%
{\renewcommand{\dashlinestretch}{30}
\begin{picture}(3309,9129)(0,-10)
\path(1047,8337)(2847,8337)(2847,7617)
	(1047,7617)(1047,8337)
\put(1947,8067){\makebox(0,0)[b]{\smash{{{\SetFigFont{12}{14.4}{\sfdefault}{\mddefault}{\updefault}build row l in}}}}}
\put(1947,7842){\makebox(0,0)[b]{\smash{{{\SetFigFont{12}{14.4}{\sfdefault}{\mddefault}{\updefault}the super-matrix}}}}}
\path(1047,7392)(2847,7392)(2847,6267)
	(1047,6267)(1047,7392)
\put(1947,7167){\makebox(0,0)[b]{\smash{{{\SetFigFont{12}{14.4}{\sfdefault}{\mddefault}{\updefault}write results to }}}}}
\put(1947,6942){\makebox(0,0)[b]{\smash{{{\SetFigFont{12}{14.4}{\sfdefault}{\mddefault}{\updefault}disk file as}}}}}
\put(1947,6717){\makebox(0,0)[b]{\smash{{{\SetFigFont{12}{14.4}{\sfdefault}{\mddefault}{\updefault}direct access}}}}}
\put(1947,6492){\makebox(0,0)[b]{\smash{{{\SetFigFont{12}{14.4}{\sfdefault}{\mddefault}{\updefault}records for bands}}}}}
\path(1047,4602)(2847,4602)(2847,4062)
	(1047,4062)(1047,4602)
\put(1947,4377){\makebox(0,0)[b]{\smash{{{\SetFigFont{12}{14.4}{\sfdefault}{\mddefault}{\updefault}recall bands from}}}}}
\put(1947,4107){\makebox(0,0)[b]{\smash{{{\SetFigFont{12}{14.4}{\sfdefault}{\mddefault}{\updefault}disk file}}}}}
\path(1047,3882)(2847,3882)(2847,3252)
	(1047,3252)(1047,3882)
\put(1947,3657){\makebox(0,0)[b]{\smash{{{\SetFigFont{12}{14.4}{\sfdefault}{\mddefault}{\updefault}build}}}}}
\put(1947,3432){\makebox(0,0)[b]{\smash{{{\SetFigFont{12}{14.4}{\sfdefault}{\mddefault}{\updefault}supermatrix}}}}}
\path(1047,1992)(2847,1992)(2847,1317)
	(1047,1317)(1047,1992)
\put(1947,1812){\makebox(0,0)[b]{\smash{{{\SetFigFont{12}{14.4}{\sfdefault}{\mddefault}{\updefault}collect the}}}}}
\put(1947,1587){\makebox(0,0)[b]{\smash{{{\SetFigFont{12}{14.4}{\sfdefault}{\mddefault}{\updefault}specific intensities}}}}}
\put(1947,1362){\makebox(0,0)[b]{\smash{{{\SetFigFont{12}{14.4}{\sfdefault}{\mddefault}{\updefault}\& build-up J's}}}}}
\path(1047,2937)(2847,2937)(2847,2172)
	(1047,2172)(1047,2937)
\put(1947,2712){\makebox(0,0)[b]{\smash{{{\SetFigFont{12}{14.4}{\sfdefault}{\mddefault}{\updefault}solve the linear}}}}}
\put(1947,2487){\makebox(0,0)[b]{\smash{{{\SetFigFont{12}{14.4}{\sfdefault}{\mddefault}{\updefault}system by}}}}}
\put(1947,2262){\makebox(0,0)[b]{\smash{{{\SetFigFont{12}{14.4}{\sfdefault}{\mddefault}{\updefault}any method}}}}}
\path(732,4962)(2937,4962)(2937,3162)
	(732,3162)(732,4962)
\path(507,5277)(3072,5277)(3072,1227)
	(507,1227)(507,5277)
\path(1947,4062)(1947,3882)
\path(1917.000,4002.000)(1947.000,3882.000)(1977.000,4002.000)
\path(1947,3162)(1947,2937)
\path(1917.000,3057.000)(1947.000,2937.000)(1977.000,3057.000)
\path(1947,2172)(1947,1992)
\path(1917.000,2112.000)(1947.000,1992.000)(1977.000,2112.000)
\put(1542,4737){\makebox(0,0)[b]{\smash{{{\SetFigFont{12}{14.4}{\sfdefault}{\mddefault}{\updefault}For all wavelength}}}}}
\put(1452,5052){\makebox(0,0)[b]{\smash{{{\SetFigFont{12}{14.4}{\sfdefault}{\mddefault}{\updefault}For all characteristics}}}}}
\path(507,1002)(3072,1002)(3072,102)
	(507,102)(507,1002)
\path(1947,1227)(1947,1002)
\path(1917.000,1122.000)(1947.000,1002.000)(1977.000,1122.000)
\path(12,5727)(3297,5727)(3297,12)
	(12,12)(12,5727)
\put(147,5502){\makebox(0,0)[lb]{\smash{{{\SetFigFont{12}{14.4}{\sfdefault}{\mddefault}{\updefault}Operator splitting iteration}}}}}
\put(1722,777){\makebox(0,0)[b]{\smash{{{\SetFigFont{12}{14.4}{\sfdefault}{\mddefault}{\updefault}Perform an OS/ALI}}}}}
\put(1722,552){\makebox(0,0)[b]{\smash{{{\SetFigFont{12}{14.4}{\sfdefault}{\mddefault}{\updefault}step with the previously}}}}}
\put(1722,327){\makebox(0,0)[b]{\smash{{{\SetFigFont{12}{14.4}{\sfdefault}{\mddefault}{\updefault}computed and saved ALO}}}}}
\path(732,8742)(2937,8742)(2937,6042)
	(732,6042)(732,8742)
\path(507,9102)(3072,9102)(3072,5907)
	(507,5907)(507,9102)
\path(1947,7617)(1947,7392)
\path(1917.000,7512.000)(1947.000,7392.000)(1977.000,7512.000)
\path(1947,5907)(1947,5727)
\path(1917.000,5847.000)(1947.000,5727.000)(1977.000,5847.000)
\put(1677,8517){\makebox(0,0)[b]{\smash{{{\SetFigFont{12}{14.4}{\sfdefault}{\mddefault}{\updefault}For all characteristics}}}}}
\put(1317,8877){\makebox(0,0)[b]{\smash{{{\SetFigFont{12}{14.4}{\sfdefault}{\mddefault}{\updefault}For all wavelength}}}}}
\end{picture}
}

\caption{\label{flow1} Flowchart for the formal solution 
process.
}
\end{figure}
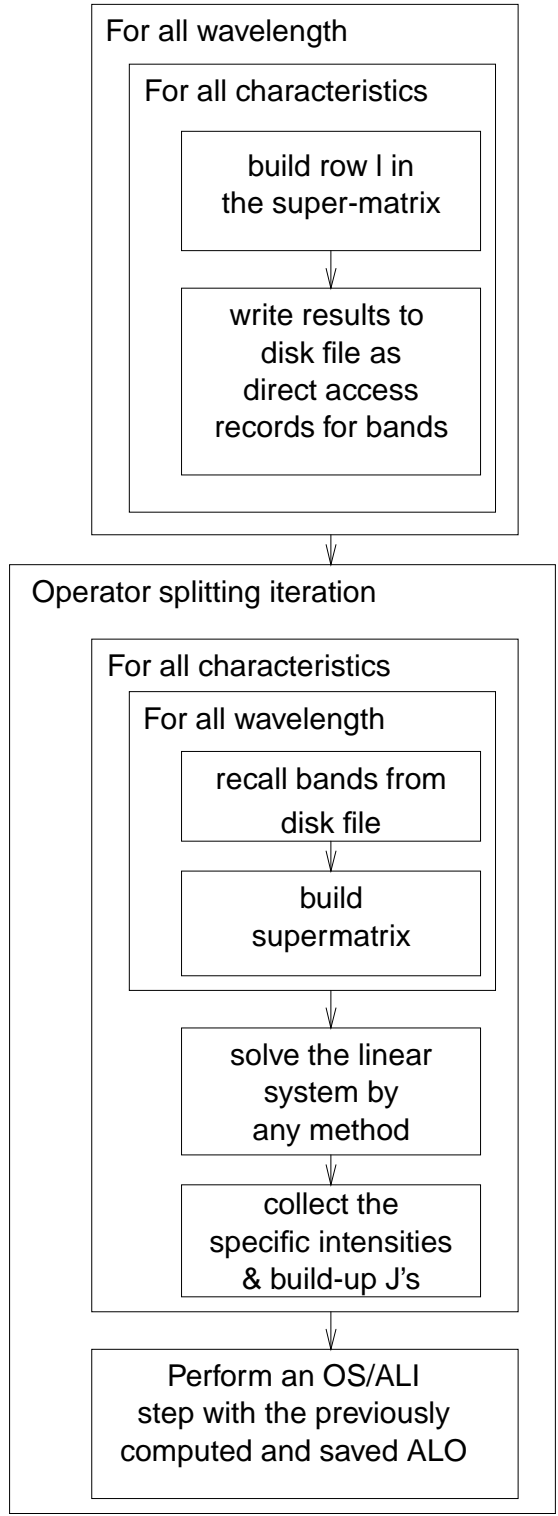

\section{Application examples}

As a first step we have implemented the method as a serial Fortran 95
program. This allows us to test the approach on problems with a
relatively small number of wavelength points.  Figure~\ref{flow1}
describes the steps involved in calculating the solution. Our basic
test problem is similar to that discussed in \citet{phhs392} and in
\citet{hb04}. We use a
spherical shell with a grey continuum opacity parameterized by a power
law in the continuum optical depth $\tstd$. The basic model parameters
are
\begin{enumerate}
\item Inner radius $r_c=10^{13}\,$cm, outer radius $\rout = 10^{15}\,$cm.
\item Minimum optical depth in the continuum $\t_{\rm std}^{\rm min} =
10^{-4}$ and maximum optical depth in the continuum $\t_{\rm std}^{\rm
max} = 10^{4}$.
\item Grey temperature structure with $\Teff=10^4$~K.
\item Outer boundary condition $I_{\rm bc}^{-} \equiv 0$ and diffusion
inner boundary condition for all wavelengths.
\item Continuum extinction $\chi_c = C/r^2$, with the constant $C$
fixed by the radius and optical depth grids.
\item Parameterized coherent \& isotropic continuum scattering by
defining
\[
\chi_c = \epsilon_c \kappa_c + (1-\epsilon_c) \sigma_c
\]
with $0\le \epsilon_c \le 1$. 
$\kappa_c$ and $\sigma_c$ are the
continuum absorption and scattering coefficients.
\item A parameterized spectral line with a rest wavelength of
$\lambda_0 = 1000\ang$ and an intrinsic width of $0.1\ang$ (equivalent
to a width of $30\kms$), the line strength is parameterized by the
ratio $\chi_l(\lambda_0)/\chi_c$, where $\chi_l$ is the line
extinction coefficient.
\item Parameterized line scattering defined analogous to the continuum
scattering with a parameter $\epsilon_l$.  We assume complete
redistribution for the line scattering.
\end{enumerate}

\subsection{Tests for static and monotonic velocity fields}

This series of tests is designed to verify the correct operation 
of the code in the monotonic velocity case where we can compare it
directly to our existing working code.

The monotonic velocity field tests assume a linear velocity law
of the form
\[
     \b(r) = \beta_0 \left(\frac{r}{\rout}\right).
\]
where $\beta_0 = v_0/c$ is the velocity at $\rout$.

\begin{figure}
\centering
\includegraphics[width=14cm,angle=90]{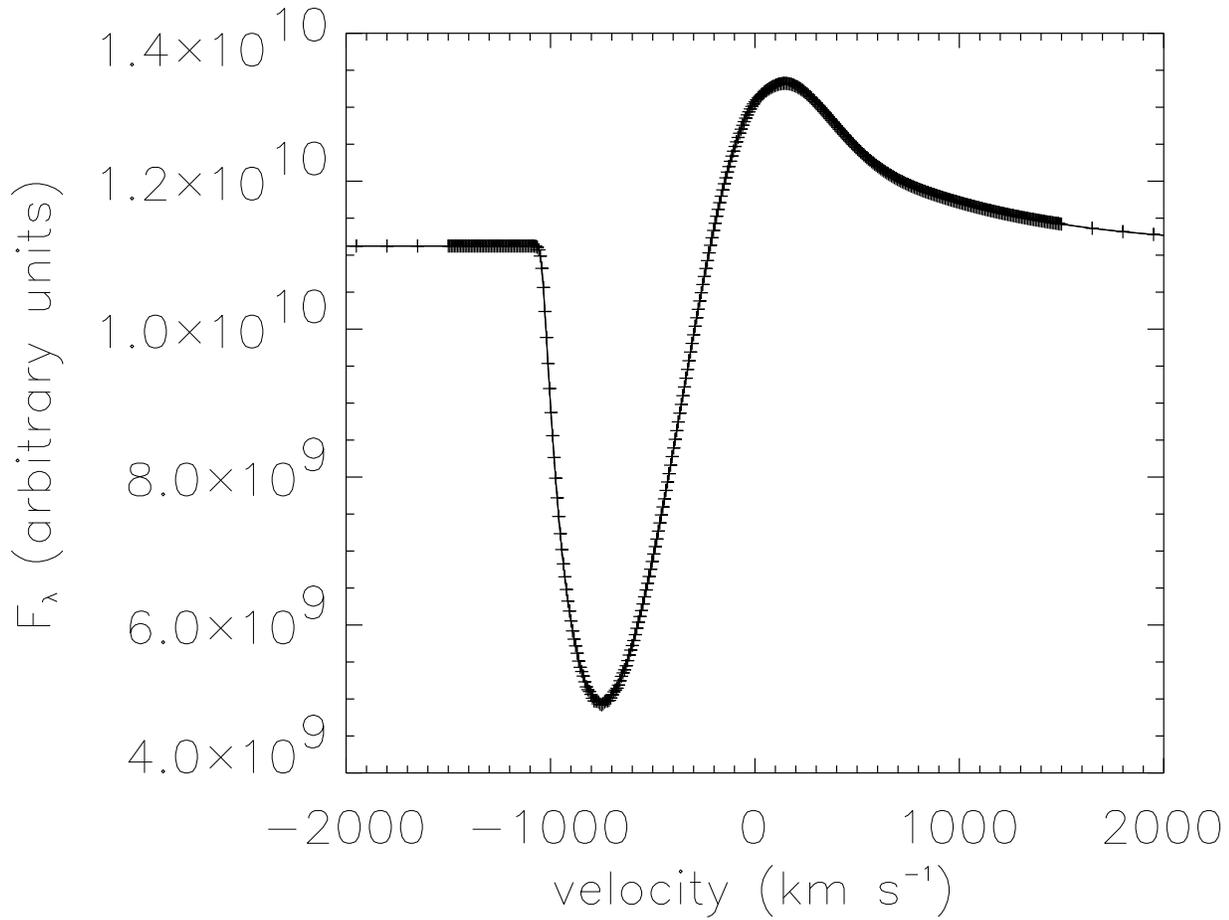}
\caption{\label{fig:homo_flux} The line profile in the observer's
frame of the homologously expanding (linear velocity law) model. The
line is the results from the monotonic velocity code \citep{phhs392,hb04}
and the points are the results from the non-monotonic code for the
case for $\epsilon_c = 0.1 \mbox{ and }
\epsilon_l = 10^{-4}$. The
results are identical.}
\end{figure}

Figure~\ref{fig:homo_flux} displays the flux transformed to the
observer's frame for a linear velocity law defined above with $v_0 =
1000~\kms$. The result is identical that produced by the purely
monotonic code to the accuracy of our models. (We require that the
scattering problem be solved to a relative accuracy of $10^{-8}$.) We
have also examined the 
moments in the co-moving frame and they too are identical to that
obtained by the purely monotonic code. Thus, the formal
solution and ALO solver and produce correct results in the regimes
where we can 
test them directly.

Table~\ref{tab:timing} shows the relative time for four different
matrix solvers: 99\% of the time is spent in solving the linear system
in the formal solution, thus the effect of varying the matrix solvers
is directly related to the code speed. Interestingly, SuperLU with the
matrix stored in standard LAPACK band form is slower than LAPACK,
where the matrix factors are saved to disk and recalled as needed;
load and store is slower than I/O! In the recall case, along a given
characteristic the ALO matrix
is fixed and thus once it has been factored it need not be factored
again, however there are too many factors to store in memory, since
the problem size is larger than available memory, particularly when
scaled to real problems requiring 300,000 wavelength points. Thus,
the factors must be written to disk and read into the appropriate arrays as
each characteristic is solved for.

A large speedup comes from storing
the SuperLU vector directly with no non-zero elements stored, even
though this involves a considerable number of ``if'' statements in the
construction loop. SuperLU can be called in a recall mode where the
factors are stored similar to the LAPACK routines, but we expect to
obtain much higher speedup by moving to SuperLU\_DIST, the parallelized
version of SuperLU and that will be discussed in a future paper.

\begin{table}
\centering
\begin{tabular}{|l|l|}
\hline
Matrix Solver&Time\\\hline
LAPACK simple&740s\\\hline
LAPACK with recall&468s\\\hline
SuperLU inefficient storage&545s\\\hline
SuperLU efficient storage&178s\\\hline
\end{tabular}
\caption{\label{tab:timing}The relative wall-clock time for a homology
  test with 
  different matrix solvers. SuperLU with efficient storage leads
  significant 
  speedup.}
\end{table}

\subsection{Non-monotonic velocity fields}

\begin{figure}
\centering
\includegraphics[width=14cm,angle=0]{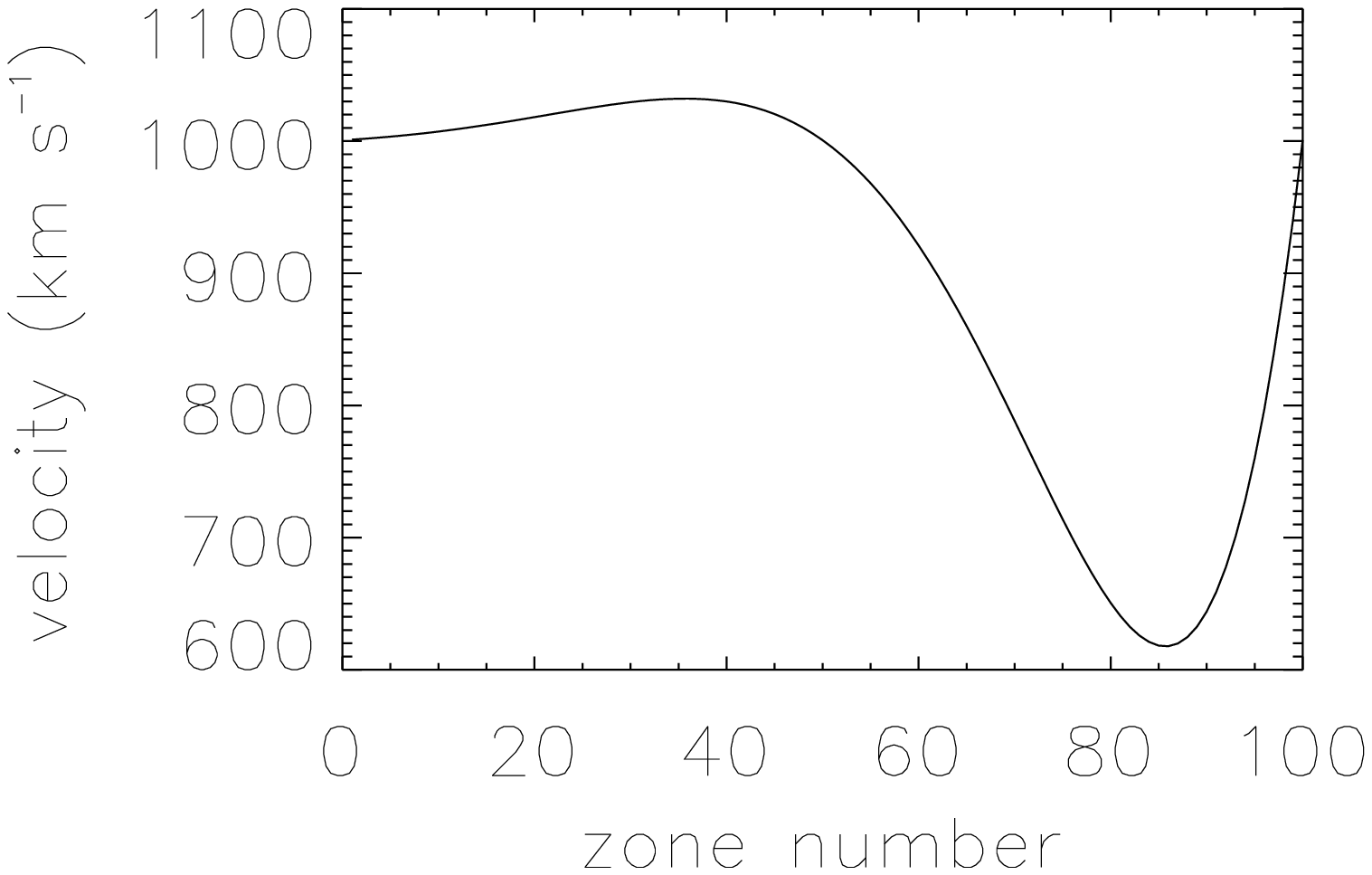}
\caption{\label{fig:sin_vel} The velocity profile of the sin wave
model. The exponential damping accounts for the damping effects of
photon viscosity.}
\end{figure}

\begin{figure}
\centering
\includegraphics[width=14cm,angle=0]{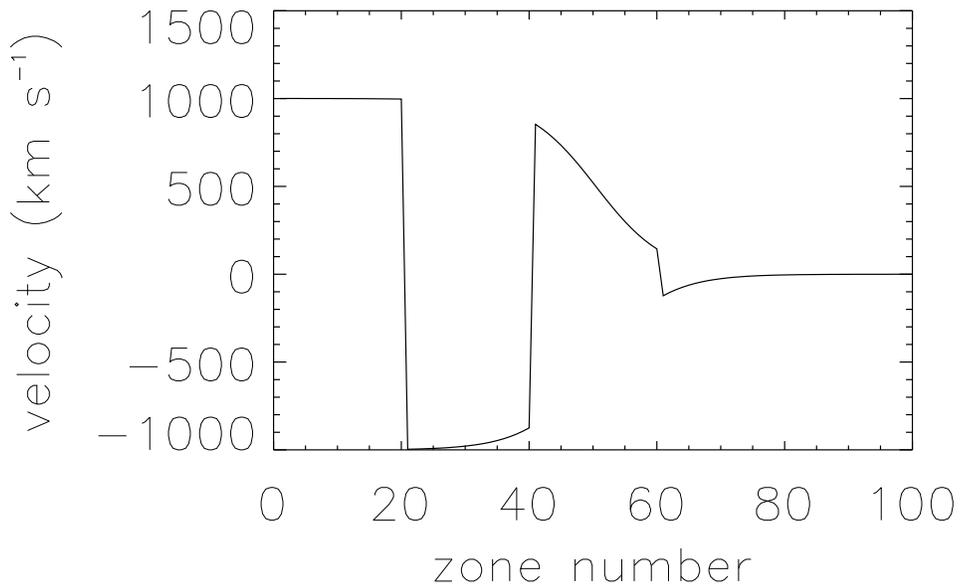}
\caption{\label{fig:shock_vel} The velocity profile of the ``shock''
model.} 
\end{figure}

\begin{figure}
\centering
\includegraphics[width=14cm,angle=90]{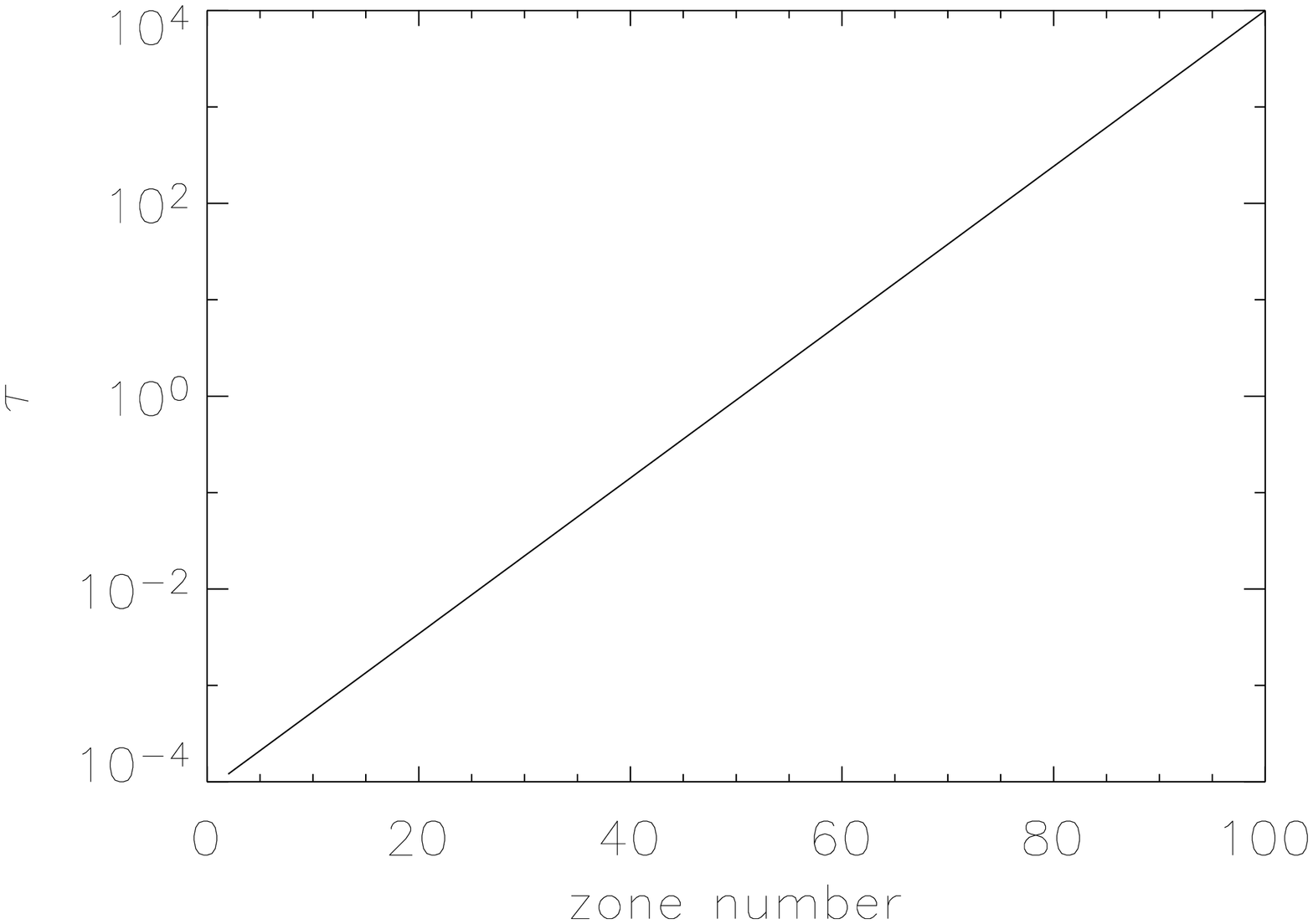}
\caption{\label{fig:tauplot} The continuum optical depth as a
function of zone number for comparision. The optical depth scale is
the same for both the sin wave model and the ``shock'' model.} 
\end{figure}

In order to test our algorithm we have assumed the velocity structures
shown in
Figs.~\ref{fig:sin_vel}--\ref{fig:shock_vel}.  The first is a
sine wave in zone number, such a structure could occur in a Mira or Cepheid
variable star.  The sine has been exponentially damped to mimic the
effects of photon viscosity, which should strongly damp the
oscillations as the material becomes optically thin
\citep{found84}.
The second, is just a illustrative set of piecewise
continuous linear velocities, that might occur in e.g., the internal
shock model of gamma-ray bursts. Figure~\ref{fig:tauplot}
shows the continuum optical depth as a function of zone number for
comparison.

\begin{figure}
\centering
\includegraphics[width=14cm,angle=90]{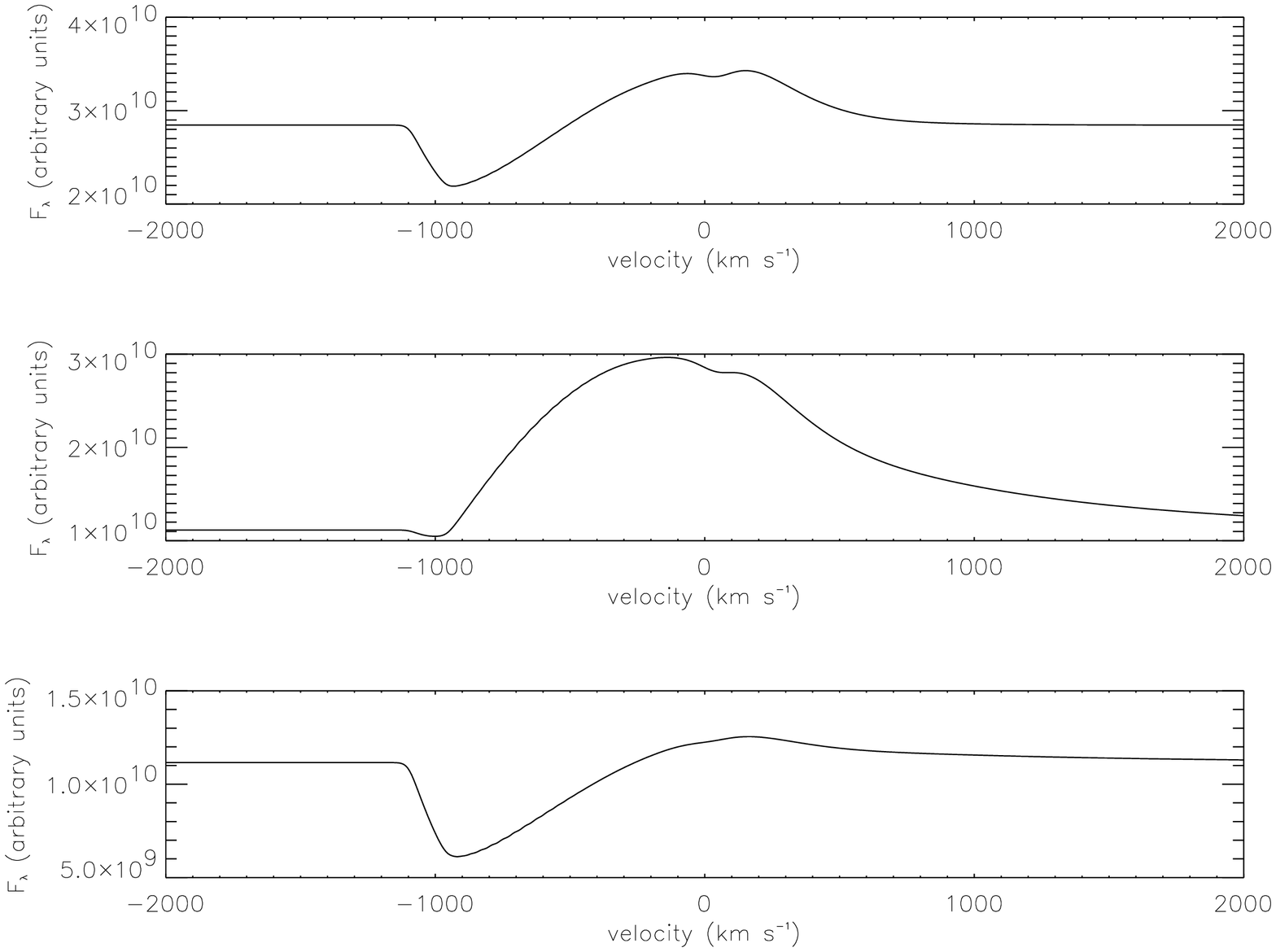}
\caption{\label{fig:sin_flux} The line profile in the observer's
frame of the sin wave model. The top panel is for the case $\epsilon_c =
\epsilon_l = 1$, the middle panel for $\epsilon_c = 0.1 \mbox{ and }
\epsilon_l = 1$, and the bottom panel for $\epsilon_c = 0.1 \mbox{ and }
\epsilon_l = 10^{-4}$.}

\end{figure}

Figure~\ref{fig:sin_flux} displays the observer's frame flux from the
sine wave model for three choices of ($\epsilon_c,\epsilon_l$). The top
panel is pure absorption in both the line and continuum, the middle
panel has a pure absorptive line in a scattering continuum, and the
bottom panel has a strongly scattering line in a scattering
continuum. At first glance the line profile seems remarkably
similar to the homologous, linear velocity law, except that there is a
small dip near zero velocity. As scattering increases, the effect of
the non-monotonic velocity law decreases. 

\begin{figure}
\centering
\includegraphics[width=14cm,angle=90]{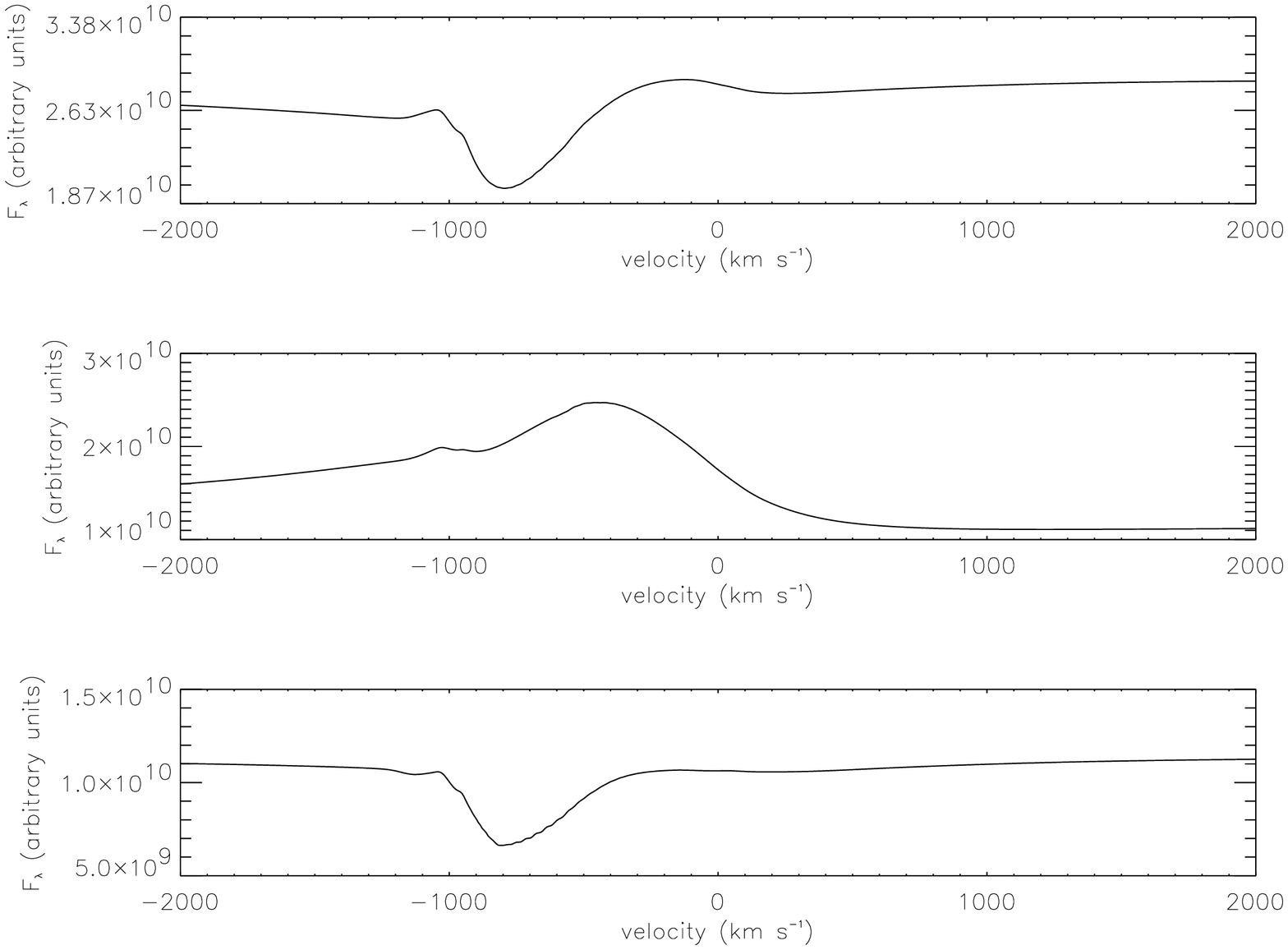}
\caption{\label{fig:shock_flux} The line profile in the observer's
frame of the ``shock'' model. The top panel is for the case $\epsilon_c =
\epsilon_l = 1$, the middle panel for $\epsilon_c = 0.1 \mbox{ and }
\epsilon_l = 1$, and the bottom panel for $\epsilon_c = 0.1 \mbox{ and }
\epsilon_l = 10^{-4}$.}

\end{figure}

Figure~\ref{fig:shock_flux} displays the observer's frame flux from
the shock wave model with the same choices for ($\epsilon_c,\epsilon_l$)
as in the sine wave model. The line profile is interesting, with
only a small ripple in the flux profile at negative
velocity. Surprisingly, this ripple does not seem to be washed out by
scattering and hence is a feature of the velocity flow.

In both cases (sine and shock) there are weak features that could be
misidentified as real weak features if only homologous flows were considered.

\section{Conclusions}

We have presented a characteristics method of solving the full 
boundary value problem that occurs in both the wavelength and spatial
dimension, using a full approximate lambda operator in space, and 
tridiagonal in wavelength. The convergence of the operator is 
slow for the wavelength-diagonal operator, much improved for the
wavelength-tridiagonal operator.  Most of the computation time is
spent in solving the linear 
system in the formal solution. This can be both sped-up and
parallelized using e.g., the SuperLU package
\citep{super_lu03}. The method we have presented is immediately useful
for studying variable stars, and can be extended to 3-D radiation
hydrodynamics and to studying other non-coherent scattering processes
such as partial redistribution, Compton scattering and neutrino
transport in Raleigh-Taylor unstable flows that occur in core collapse
supernovae.

\begin{acknowledgements}
We thank Sherry Xiaoye for helpful tutelage in the use of Super\_LU.
This work was supported in part by by NASA grant
NAG5-3505, NSF grants AST-0204771 and AST-0307323, an IBM SUR
grant to the University of Oklahoma and by 
NASA grants NAG 5-8425 and NAG 5-3619 to
the University of Georgia. PHH was 
supported in part by the P\^ole Scientifique de Mod\'elisation
Num\'erique at ENS-Lyon. Some of the calculations presented here were
performed at the San Diego Supercomputer Center (SDSC), supported by
the NSF, at the National Energy Research Supercomputer Center
(NERSC), supported by the U.S. DOE, and at the H\"ochstleistungs
Rechenzentrum Nord (HLRN).  We thank all these institutions 
for a generous allocation of computer time.
\end{acknowledgements}

\clearpage

\bibliography{refs,baron,rte,crossrefs}

\begin{thebibliography}{20}
\expandafter\ifx\csname natexlab\endcsname\relax\def\natexlab#1{#1}\fi

\bibitem[{Auer \& Blerkom(1972)}]{avb72}
Auer, L.~H. \& Blerkom, D.~V. 1972, ApJ, 178, 175

\bibitem[{Cannon(1973)}]{cannon73}
Cannon, C.~J. 1973, JQSRT, 13, 627

\bibitem[{Caroff {et~al.}(1972)Caroff, Noerdlinger, \& Scargle}]{cns72}
Caroff, L.~J., Noerdlinger, P.~D., \& Scargle, J.~D. 1972, ApJ, 176, 439

\bibitem[{Castor(1970)}]{castor70}
Castor, J.~I. 1970, MNRAS, 149, 111

\bibitem[{Golub \& Loan(1989)}]{golub89:_matrix}
Golub, G.~H. \& Loan, C.~V. 1989, Matrix computations (Baltimore: Johns Hopkins
  University Press)

\bibitem[{Hamann(1987)}]{hamann87}
Hamann, W.-R. 1987, in Numerical Radiative Transfer, ed. W.~Kalkofen
  (Cambridge: Cambridge Univ. Press), 35

\bibitem[{Hauschildt(1992)}]{phhs392}
Hauschildt, P.~H. 1992, JQSRT, 47, 433

\bibitem[{Hauschildt \& Baron(2004)}]{hb04}
Hauschildt, P.~H. \& Baron, E. 2004, A\&A, 417, 317

\bibitem[{Hauschildt {et~al.}(1994)Hauschildt, St{\"o}rzer, \& Baron}]{hsb94}
Hauschildt, P.~H., St{\"o}rzer, H., \& Baron, E. 1994, JQSRT, 51, 875

\bibitem[{Hauschildt \& Wehrse(1991)}]{phhrw91}
Hauschildt, P.~H. \& Wehrse, R. 1991, JQSRT, 46, 81

\bibitem[{Kalkofen(1987)}]{kalk87}
Kalkofen, W., ed. 1987 (Cambridge: Cambridge Univ. Press)

\bibitem[{Magnan(1970)}]{magnan70}
Magnan, C. 1970, JQSRT, 10, 1

\bibitem[{Mihalas {et~al.}(1976)Mihalas, Kunasz, \& Hummer}]{mkh76b}
Mihalas, D., Kunasz, P., \& Hummer, D. 1976, ApJ, 206, 515

\bibitem[{Mihalas \& Mihalas(1984)}]{found84}
Mihalas, D. \& Mihalas, B.~W. 1984, Foundations of Radiation Hydrodynamics
  (Oxford: Oxford University)

\bibitem[{Olson {et~al.}(1987)Olson, Auer, \& Buchler}]{OAB}
Olson, G.~L., Auer, L.~H., \& Buchler, J.~R. 1987, JQSRT, 38, 431

\bibitem[{Olson \& Kunasz(1987)}]{ok87}
Olson, G.~L. \& Kunasz, P.~B. 1987, JQSRT, 38, 325

\bibitem[{Scharmer(1984)}]{scharmer84}
Scharmer, G.~B. 1984, in Methods in Radiative Transfer, ed. W.~Kalkofen
  (Cambridge: Cambridge Univ. Press), 173

\bibitem[{Werner(1987)}]{werner87}
Werner, K. 1987, in Numerical Radiative Transfer, ed. W.~Kalkofen (Cambridge:
  Cambridge Univ. Press), 67

\bibitem[{Xiaoye \& Demmel(2003)}]{super_lu03}
Xiaoye, S.~L. \& Demmel, J.~W. 2003, AM Transaction on Mathematical Software,
  29, 110

\bibitem[{Zurm{\"u}hl \& Falk(1986)}]{matrizen}
Zurm{\"u}hl, R. \& Falk, S. 1986, Matrizen und ihre Anwendungen, 5th edn.,
  Vol.~2 (Berlin: Springer-Verlag)

\end{thebibliography}

\end{document}